# Observation of Non-Hermitian Skin Effect in Thermal Diffusion

Yun-Kai Liu [1,*], Pei-Chao Cao [2,3,4,5,*], Minghong Qi[2,3,4,5], Qiang-Kai-Lai Huang[2,3,4,5], Yu-Gui Peng[1,†], Ying Li[2,3,4,5,†], Xue-Feng Zhu[1,†]

[1]*School of Physics and Innovation Institute, Huazhong University of Science and Technology, Wuhan 430074, China*

[2]*Interdisciplinary Center for Quantum Information, State Key Laboratory of Extreme Photonics and Instrumentation, ZJU-Hangzhou Global Scientific and Technological Innovation Center, Zhejiang University, Hangzhou 310027, China.*

[3]*International Joint Innovation Center, The Electromagnetics Academy at Zhejiang University, Zhejiang University, Haining 314400, China*

[4]*Key Lab. of Advanced Micro/Nano Electronic Devices & Smart Systems of Zhejiang, Jinhua Institute of Zhejiang University, Zhejiang University, Jinhua 321099, China*

[5]*Shaoxing Institute of Zhejiang University, Zhejiang University, Shaoxing 312000, China*

[*] These authors contributed equally to this work.

[†]Corresponding authors: ygpeng@hust.edu.cn (Y. -G. P.); eleying@zju.edu.cn (Y. L.); xfzhu@hust.edu.cn (X.-F. Z.)

The paradigm shift of the Hermitian systems into the non-Hermitian regime profoundly modifies the inherent topological property, leading to various unprecedented effects such as the non-Hermitian skin effect (NHSE). In the past decade, the NHSE effect has been demonstrated in quantum, optical and acoustic systems. Besides in those non-Hermitian wave systems, the NHSE in diffusive systems has not yet been explicitly demonstrated, despite recent abundant advances in the study of topological thermal diffusion. Here we first design a thermal diffusion lattice based on a modified Su-Schrieffer-Heeger model which enables the observation of diffusive NHSE. In the proposed model, the periodic heat exchange rate among adjacent unit cells and the asymmetric temperature field coupling inside unit cells can be judiciously realized by appropriate configurations of structural parameters of unit cells. The transient concentration feature of temperature field on the boundary regardless of initial excitation conditions can be clearly observed, indicating the occurrence of transient thermal skin effect. Nonetheless, we experimentally demonstrated the NHSE and verified the remarkable robustness against various defects. Our work provides a platform for exploration of non-Hermitian physics in the diffusive systems, which has important applications in efficient heat collection, highly sensitive thermal sensing and others.

After the birth of topological insulators in condensed matter physics, the study of topological phases of matter has garnered significant attentions across the different fields of physics[1-3]. In conventional topological band theory, the existence of robust boundary state is determined by the topological invariants defined by the Bloch Hamiltonian. The systematic Hamiltonians are considered to be either Hermitian or anti-Hermitian. Introducing the boundaries will give rise to nonaccidental appearance of edge states without altering the bulk properties[4-6]. However, for open systems, such as classical wave systems of gain and/or loss modulation[7-15] and interacting electronic systems[16-18], the description of Hermiticity will break down and the non-Hermitian Hamiltonians provide an appropriate approximation to describe the intrinsic properties. In stark contrast to the Hermitian systems, the Hamiltonians of non-Hermitian systems normally have complex eigenvalues with non-orthogonal eigenmodes[19], leading to the fruitful discoveries such as exceptional points and parity-time symmetry breaking[20-24]. Recently, the topological phase transition was investigated in the non-Hermitian systems with numerous unique phenomena and interesting functionalities, thus making the non-Hermitian topological physics a thriving research area[25-30].

In the past years, the intriguing non-Hermitian skin effect (NHSE) has been discovered in open systems, whereby all the eigenmodes decay exponentially and localize at the open boundaries[31-38]. This feature indicates interesting physics of NHSE in classical and quantum systems. The NHSE reflects the existence of peculiar gap topology that is unique in non-Hermitian systems[39-41] and alters the bulk-boundary correspondence in the conventional wisdom[31, 33, 42], expanding the research scope in topological phases of matter. Recently, NHSE was discovered to have vibrant applications in different areas such as enhanced sensing[43,44], topological lasers[45], and light funneling[32] owing to its unique properties. The experimental verifications of NHSE hitherto have been achieved in various systems, including the phononic crystals,[46,47] synthetic lattices[32], and electric ciruits[48].

However, NHSE have not yet been experimentally observed in diffusive systems. Unlike those wave systems governed by Hermitian Hamiltonians with real eigenvalues, the diffusive systems are dissipative and the coupling coefficients between separate meta-atoms are purely imaginary, providing a natural platform for studying the non-Hermitian physics[21,49,50]. For example, the researchers have demonstrated many interesting phenomena in the diffusive-convective systems, such as the anti-parity-time symmetric phase transition at an exceptional point[23,51], the topological heat localization[52] and chiral heat transport[50]. One recent theoretical proposal has shown the possibility of NHSE in thermal diffusion[53], which renders an automatic concentration of the temperature field towards the boundary regardless of initial heat excitation. Due to the existing challenges in realizing precise modulation of material parameters, such as thermal conductivity and mass density, the experimental demonstration of diffusive NHSE has remained elusive hitherto.

In this article, we design a one-dimensional (1D) thermal diffusion lattice based on the modified Su-Schrieffer-Heeger (SSH) model and report the first experimental demonstration of NHSE in diffusive systems for the transient directional temperature field concentration. The required modulation of material parameters is implemented by the judicious configuration of structural parameters of site rods and coupling sticks, so as to flexibly manipulate temperature field couplings between the neighboring sites. By tailoring the asymmetric couplings inside each unit cell, we find that the thermal system can change from anti-Hermitian to non-Hermitian. When the asymmetry coupling factor deviates from unitary, all eigenmodes take the form of localized skin modes on the boundary. Here we first verify the theoretical prediction of diffusive NHSE in numerical simulations with effective material parameters. The temperature field is proved to concentrate to the designated boundary in a transient process, regardless of the initial excitations. Then we fabricated the sample with the structural parameters satisfying the coupling parameters and observed the temperature field evolution to concentrate towards the target boundary in a vacuum chamber. We further demonstrated the robustness of diffusive

NHSE against different defects. Our work provides a platform for a distinctive type of thermal metamaterials and functional thermo-devices based on asymmetric temperature field couplings. The experimental realization of transient diffusive NHSE opens the door for the realization of high-precision thermal sensing and robust heat harvesting.

**Results**

**NHSE based on asymmetric coupling dimer unit-cells**

In nature, thermal diffusion intrinsically enables heat to spontaneously spread and the temperature field will distribute evenly in materials according to the zeroth law of thermodynamics, as shown in Fig. 1a. However, in the transient process, it is possible to break the symmetry via thermal metamaterials to manipulate the temperature field preferentially concentrating at the target boundary, as sketched in Fig. 1a, denoting the proposed diffusive skin effect. The key recipe for the diffusive skin effect here is the non-Hermitian asymmetric coupling in the unit-cell, for instance, a coupled cavity dimer. As shown in Fig. 1b, a simple cavity dimer model is presented, where the temperature field coupling can be asymmetric, with the heat exchange efficiencies between $A$ and $B$ cavities ($D_{AB}$) or between $B$ and $A$ cavities ($D_{BA}$) being $D/a$ and $Da$, respectively. Here a non-zero parameter $a$ is utilized in combination with the heat transfer efficiency $D$ to construct the asymmetric heat exchange. According to the Fourier's law in heat conduction, the coupling equations can be written as

$$\begin{aligned} \frac{\partial T_A}{\partial t} &= D_{BA}(T_B - T_A) \\ \frac{\partial T_B}{\partial t} &= D_{AB}(T_A - T_B) \end{aligned}, \quad (1)$$

where $T_A$ and $T_B$ are the temperature fields in $A$ and $B$ cavities, and $t$ denotes the time.

We consider that the Eq. (1) has a wavelike solution, which takes the form of $T = Ae^{-i\omega t} + T_0$, where $A$ and $\omega$ are the amplitude and complex frequency of the temperature field, and $T_0$ is a constant[21,51,54]. By using the wavelike solution, Eq. (1) can be transformed into an eigenvalue problem with the effective Hamiltonian $H$,

$$H = -i \begin{pmatrix} Da & -Da \\ -\frac{D}{a} & \frac{D}{a} \end{pmatrix}. \tag{2}$$

Note that this diffusive Hamiltonian is purely imaginary, very different from the one in wave systems with real Hamiltonians. For Eq. (2), the eigenvalues can be solved as $\omega_1 = 0$ and $\omega_2 = -i\frac{Da^2+D}{a}$, for which the corresponding eigenstates are $u_1 = (1,1)^T$ and $u_2 = (-a^2, 1)^T$. For $\omega_1 = 0$, it can be found that the temperature field is symmetrical in the cavity dimer unit cell. However, there also exists an eigenvector $u_2$ that satisfies an asymmetric temperature field coupling in a dimer with an asymmetrically coupling factor $a$. When the heat exchange efficiencies $\frac{D}{a} \neq Da$, we will obtain the asymmetric temperature field distributions in the cavity dimer unit cell, as shown in Fig. 1b. The full wave simulations in Fig. 1c further present that the degree of asymmetry for temperature field coupling increases as the factor $a$ becomes smaller.

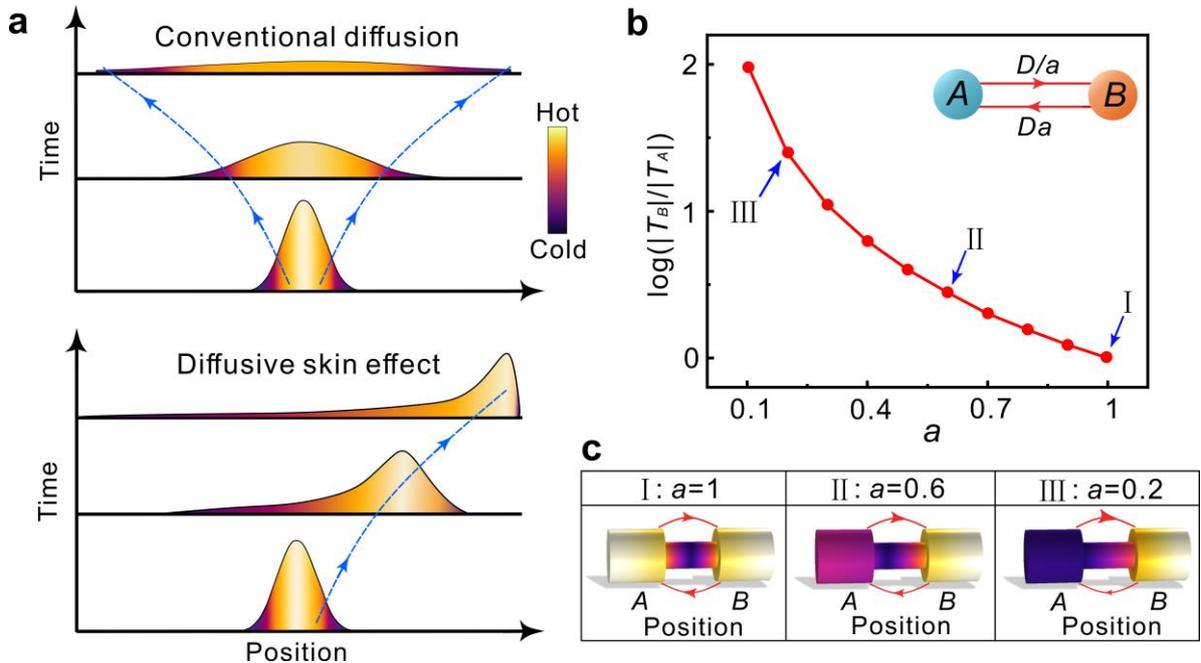

**Fig. 1 | Diffusive skin effect and asymmetric coupling. a** The concept of transient diffusive skin effect in comparison with the conventional heat diffusion. **b** The temperature field ratios versus an asymmetric factor in a basic unit cell for realizing the transient diffusive NHSE, where

a coupled cavity dimer with the judicious design can well support the asymmetric coupling of temperature fields. **c** The simulated temperature field distributions in the coupled cavity dimers with different asymmetric factors.

**Topology and skin effect in the diffusive SSH model**

The cavity dimer with asymmetric temperature field coupling inside can further be aggregated as a 1D chain to form a two-band Su-Schrieffer-Heeger (SSH) model, as illustrated in Fig. 2a, where the NHSE can be readily observed. Specifically, the couplings from left to right are $\frac{D_2}{a}$ for the intra-cell coupling coefficient and $D_1$ for the inter-cell coupling coefficient, as indicated by the red and blue arrows in the upper part of the lattice. In contrast, the couplings from right to left as marked by the arrows in the lower part are $D_2 a$ for the intra-cell coupling coefficient and $D_1$ for the inter-cell coupling coefficient, respectively. Following the Fourier's law, the temperature field evolution at each lattice site can be written as

$$\begin{aligned} \partial_t T_{A_{N-1}} &= D_1\big(T_{B_{N-1}} - T_{A_N}\big) + D_2 a\big(T_{B_N} - T_{A_N}\big) \\ \partial_t T_{B_N} &= \frac{D_2}{a}\big(T_{A_N} - T_{B_N}\big) + D_1\big(T_{A_{N+1}} - T_{B_N}\big) \end{aligned}, \quad (3)$$

where $T_{A_N}(T_{B_N})$ is the temperature field at site $A(B)$ in the $N$th period. Then applying the periodic boundary condition (PBC), the corresponding Hamiltonian in the reciprocal space can be obtained as

$$H_{\text{SSH}}(k) = -i \begin{pmatrix} D_2 a + D_1 & -D_2 a - D_1 e^{-ik} \\ -\frac{D_2}{a} - D_1 e^{ik} & \frac{D_2}{a} + D_1 \end{pmatrix}, \quad (4)$$

where $k$ is the wave number and only the nearest coupling is considered for the tight binding model. The eigenvalues solved by Eq. (4) have the following form

$$\omega(k) = -i\left[\frac{2D_1 + D_2 a + \frac{D_2}{a}}{2} \pm \frac{\sqrt{\left(2D_1 + D_2 a + \frac{D_2}{a}\right)^2 - 4\left[D_1 D_2 a(1 - \cos k - i\sin k) + \frac{D_1 D_2}{a}(1 - \cos k + i\sin k)\right]}}{2}\right]. \quad (5)$$

To facilitate analysis, we show the energy spectra for $D_1 = D_2 = 0.0072$ and different values of $a$ in Figs. 2b-2d. In each figure, the blue solid line corresponds to PBC, and the orange dots correspond to OBC for the lattice with 10 unit-cells. For the case of $a = 1$ in Fig. 2b, the intra-cell and inter-cell coupling coefficients are symmetrical as $\frac{D_2}{a} = D_2 a = D_1$. The governing Hamiltonian describing the temperature field evolution is anti-Hermitian, thereby resulting in a purely imaginary energy spectrum $\omega$. The eigenfield distributions for bulk states manifest the extended profiles (see Supplementary Note 1). When we further set the asymmetry factor $a \neq 1$ and generate a lattice with asymmetric intra-cell coupling coefficients of $\frac{D_2}{a}$ and $D_2 a$. In this situation, the NHSE model becomes non-Hermitian, where the eigenvalues under PBC have nonzero real parts $\text{Re}(\omega)$ and a band gap emerges, as shown in Fig. 2c, 2d. The energy spectra in diffusive and wave systems are very different from each other, for which the band gaps are open and closed at $k = \pm\pi$ for $a \neq 1$, respectively (see Supplementary Note 2). However, this major difference does not affect the appearance of NHSE in both systems. In diffusive systems, we note that the eigenvalues of OBC are all surrounded by the eigenvalue loops of PBC, indicating the breakdown of bulk-boundary correspondence and revealing the non-Bloch bulk-boundary correspondence of diffusive system (see Supplementary Note 1). The enclosing loop in Figs. 2c, 2d implies the existence of NHSE. To describe the looping property of this complex plane spectrum, one can define a quantized winding number of $\omega(k)$[39, 44]

$$W = \int_{-\pi}^{\pi} \frac{1}{2\pi i} \frac{\partial \omega(k)/\partial k}{\omega(k) - \omega_0} dk, \quad (6)$$

where $\omega(k)$ is the eigenspectrum over the Brillouin zone with PBC, and $\omega_0$ is reference point of frequency given arbitrarily. If there exists NHSE in the system and the skin mode is localized at the right (or left) boundary, then $W$ is calculated to be $W = -1$ (or $+1$) at the reference frequency of $\omega_0$. In our case, Eq. (6) gives out $W = -1$, which means that eigenstates exhibit the skin mode at the rightside boundary (see Supplementary Note 3). In addition, the area size of closed loops reflects the strength of skin effect in accumulating the temperature fields. In Fig. 2e, we present the eigenfield distributions of eigenmodes at the OBC for $a = 0.5$. The result clearly shows that the corresponding temperature fields of eigenmodes are finely localized at one boundary (or nearby site 20), representing a typical feature of NHSE. The position of temperature localization can also be verified numerically by solving the Eq. (6) to check out the sign of winding numbers.

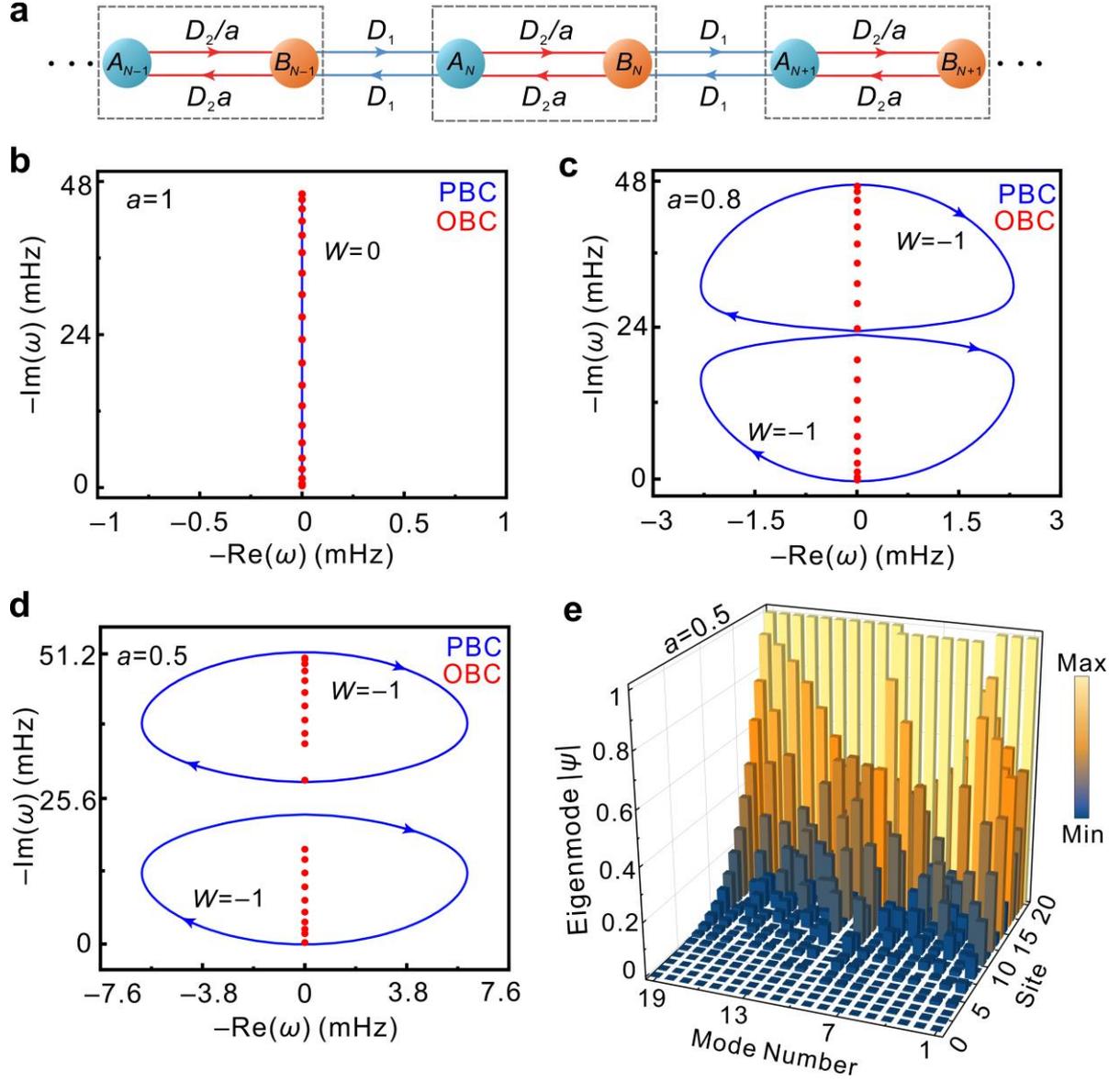

**Fig. 2 | 1D NHSE in the diffusive SSH model. a** The analytical model of diffusive NHSE with the asymmetric intra-cell coupling coefficients $\frac{D_2}{a}$ and $D_2 a$ as well as the symmetric inter-cell coupling coefficient of $D_1$. **b-d** The calculated complex eigenfrequency spectra with trivial and nontrivial winding numbers for the periodic boundary condition (PBC) as well as the open boundary condition (OBC), when $a = 1, 0.8, 0.5$, respectively. **e** The calculated eigenfield

distributions of the eigenstates at $a = 0.5$ based on the governed Hamiltonian, showing the non-Hermitian skin effect that is predicted in the analytic solution.

**Simulation of transient diffusive NHSE**

In the implementation of transient diffusive NHSE, we first construct a 1D thermal lattice for simulation, which consists of periodical rods connected by sticks as shown in Fig. 3a. The cylindrical rods have the radius $R = 8$ mm and length $b = 16$ mm, with the volume $V = \pi R^2 b$. The coupling sticks have the radius $r = 4$ mm and length $d = 16$ mm, with the cross-section area $S = \pi r^2$. For the thermal lattice, which contains $N$ periods and $2N$ sites, the lattice constant $l = 2(b + d)$. The volume specific heat capacities $\rho c_{p_n}$ and $\rho c_{p_{n+1}}$ of two adjacent rods numbered as $n$ and $n + 1$ influence each other through a connecting stick with the thermal conductivity being $\kappa_n$. Here we define an initial coupling coefficient $D_0 = \frac{\kappa_0 S}{\rho c_{p_0} V d}$ and let the effective volume specific heat capacity and thermal conductivity ($\rho c_{p_n}$ and $\kappa_n$) be set based on the form given in Supplementary Note 4

$$\rho c_{p_n} = \begin{cases} a^{n-1}\rho c_{p_0}, n = 1, 3, 5, \cdots \\ a^n \rho c_{p_0}, n = 2, 4, 6, \cdots \end{cases}, \kappa_n = a^n \kappa_0, n = 1,2,3,\cdots. \quad (7)$$

In this case, the asymmetric intra-cell coupling coefficients can be expressed as $\frac{1}{a}\frac{\kappa_0 S}{\rho c_{p_0} V d}$ and $a\frac{\kappa_0 S}{\rho c_{p_0} V d}$, where the inter-cell coupling coefficient is $\frac{\kappa_0 S}{\rho c_{p_0} V d}$. In Fig. 3, we simulate the 1D thermal lattice consisting of 10 unit cells and with the asymmetric factor of $a = 0.5$ by using a commercial finite element solver (COMSOL Multiphysics 6.0). The theoretical and simulation results of eigenvalues are shown in Fig. 3b, represented by the red and blue dots, respectively (see Supplementary Note 4). Note that the lower band and the band gap agree well with each other. However, the higher band in simulations deviates much from the theoretically calculated one. The discrepancy primarily roots in the reason that as $-\mathrm{Im}(\omega)$ decreases, the tight-binding model of 1D thermal lattice is no longer appropriate to be used for the description of continuous heat diffusion process. We further simulate the eigenmode profiles in Fig. 3c, which clearly present the temperature field concentration towards the target boundary, which is consistent with theoretical analyses above.

In Figs. 3d and 3e, we conduct the transient simulation of temperature field evolution to verify the existence of transient diffusive NHSE. In the first case of Fig. 3d, we impose a heat source to the 7th cylindrical rod and heat it to 343.15 K, while other site rods maintain the room temperature of 293.15 K. The freeze-frames of temperature fields at six different times ($t = 1, 100, 200, 300, 400, 500$ s) are shown in Fig. 3d. For convenience, the temperature field is normalized by $T_{norm} = (T - T_L)/(T_{max} - T_{min})$, where $T_L$ is the temperature at the leftmost site and $T_{max}$ ($T_{min}$) is the maximum (minimum) temperature on the thermal lattice. The result shows that the region where the maximum temperature field locates keeps on moving towards the rightside boundary in a transient process under the drive of asymmetric temperature field coupling. At around $t = 500$ s, the temperature field concentrates at the right boundary of the thermal lattice. In the second case of Fig. 3e, we further impose a random heat source on the thermal lattice. For example, we heat the 5th, 9th and 13th site rods up to 338.15 K, 333.15 K, and 318.15 K, while we cool the 4th, 6th, and 11th site rods down to 323.15 K, 333.15 K, and 333.15 K, with the remaining site rods at the room temperature of 293.15 K. The transient temperature field evolution corresponding to different times is presented in Fig. 3e. In close resemblance to the result in Fig. 3d, the temperature field in Fig. 3e evolves to concentrate at the rightside boundary of the thermal lattice ($t = 500$ s). The numerical simulation indicates that the thermal lattice we proposed in this article can be used as a platform for realizing the transient diffusive NHSE, for which temperature field will be localized at the targeted boundary over time regardless of the initial heat excitation conditions.

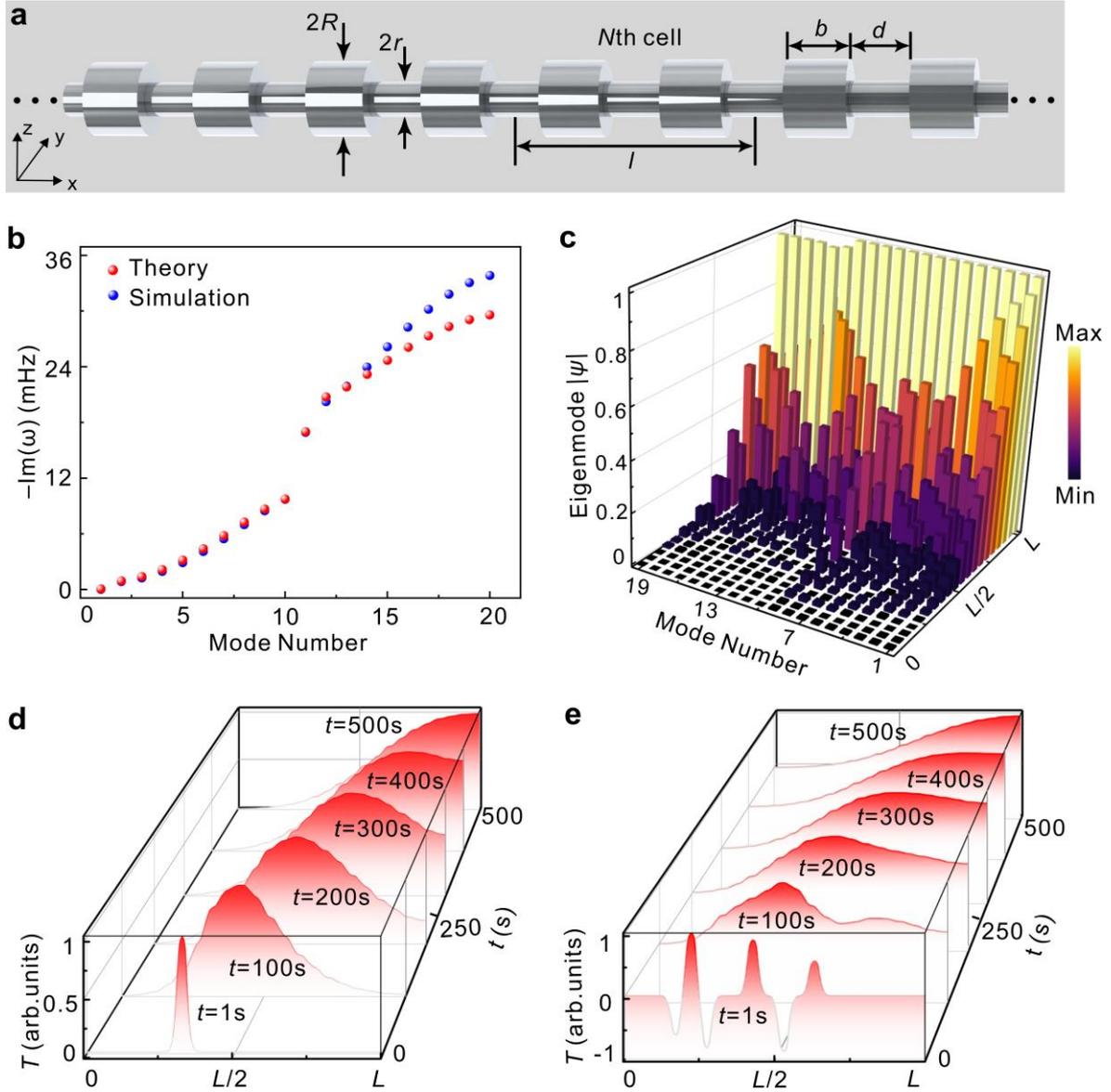

**Fig. 3 | The designed thermal lattice for simulating transient diffusive NHSE. a** A schematic of 1D cavity chain for simulating the diffusive NHSE. The geometrical parameters are marked, which are closely related to the effective material parameters. **b** The calculated eigenvalues in a finite thermal lattice (10 unit cells) for theoretical analysis and simulation. **c** The eigenfield distributions of the eigenmodes in numerical simulations. **d, e** Evolution of temperature fields with time $t$ under different initial excitations, verifying the existence of the transient diffusive NHSE.

**Experimental demonstration of transient diffusive NHSE**

To experimentally observe diffusive NHSE, the main challenge roots in the realization of effective volume specific heat capacity and thermal conductivity given in Eq. (7). Here we overcome the challenge by resorting to the thermal metamaterials and utilizing the structural parameters to engineer the effective coupling coefficients. For example, the designed sample for experiments, as shown in Fig. 4a, is made of Aluminium Alloy (AlSi$_{10}$Mg) with thermal conductivity $\kappa = 180$ W/m $\cdot$ K, density $\rho = 2700$ kg/m$^3$, and heat capacity $c = 900$ J/kg $\cdot$ K. The length of site rod is equal to the length of coupling stick with $b = d = 16$ mm. The volume of cylindrical site rod is $V_n = \pi R_n^2 b$ and the cross-section area of cylindrical coupling stick is $S_n = \pi r_n^2$. Therefore, the coupling coefficients between site rods $n$ and $n+1$ take the form of $D_{n,n+1} = \frac{\kappa S_n}{\rho c d V_{n+1}}$ for the left-to-right coupling and $D_{n+1,n} = \frac{\kappa S_n}{\rho c d V_n}$ for the right-to-left coupling, respectively. In order to make the adjacent site coupling of the sample meet the requirements given in Eq. (3), we construct the thermal lattice with the structural parameters following

$$R_n = \begin{cases} a^{\frac{n-1}{2}} R_0, n = 1, 3, 5, \cdots \\ a^{\frac{n}{2}} r_0, n = 2, 4, 6, \cdots \end{cases}, r_n = a^{\frac{n}{2}} r_0, n = 1, 2, 3, \cdots, \qquad (8)$$

where the parameters $R_0$ and $r_0$ are set to 8 mm and 4 mm, respectively. In Fig. 4b, we show the structural parameter settings of each site rod (blue bars) and coupling stick (red bars) of the fabricated thermal lattice. In our case, the factor $a$ is set to be 0.7 for the convenience of 3D metal printing. The experimental measurement was conducted in a vacuum chamber, where we used a resistance wire as the heat source and the silicone grease was covered to improve the contact between the resistance wire and the sample for better heat transfer. A current was loaded to the resistance wire to generate a large amount of heat in a short time, which was transferred to the target site rod as marked by the red asterisk in Fig. 4a. More details of the experiments are appended in Supplementary Note 5. We present the starting and final states of temperature field distributions in Fig. 4c. The experimental results are highly consistent with the simulation ones, except for the heated region where the resistance wires and silicone grease cover. Due to the very different emissivity, the heated region marked by a gray bar has an abrupt temperature variation in the thermograph. However, the discrepancy at the heated region does not affect the overall evolution of temperature field. For example, at $t = 8$ s, the temperature field is localized in the vicinity of heated region, as shown by the red (experiment) and blue (simulation) curves. After the evolution at $t = 120$ s, we observed a significant change of temperature field, which was concentrated at the rightside boundary, as indicated by the green (experiment) and purple (simulation) curves. In Fig. 4d, we show three freeze-frames of thermographs taken at $t = 8, 60, 120$ s. The direction of temperature field flows is marked by a blue dashed arrow line. The experimental results clearly show that driven by the asymmetric couplings, the temperature field gradually concentrates towards the rightside end, confirming the transient diffusive NHSE in the fabricated thermal lattice.

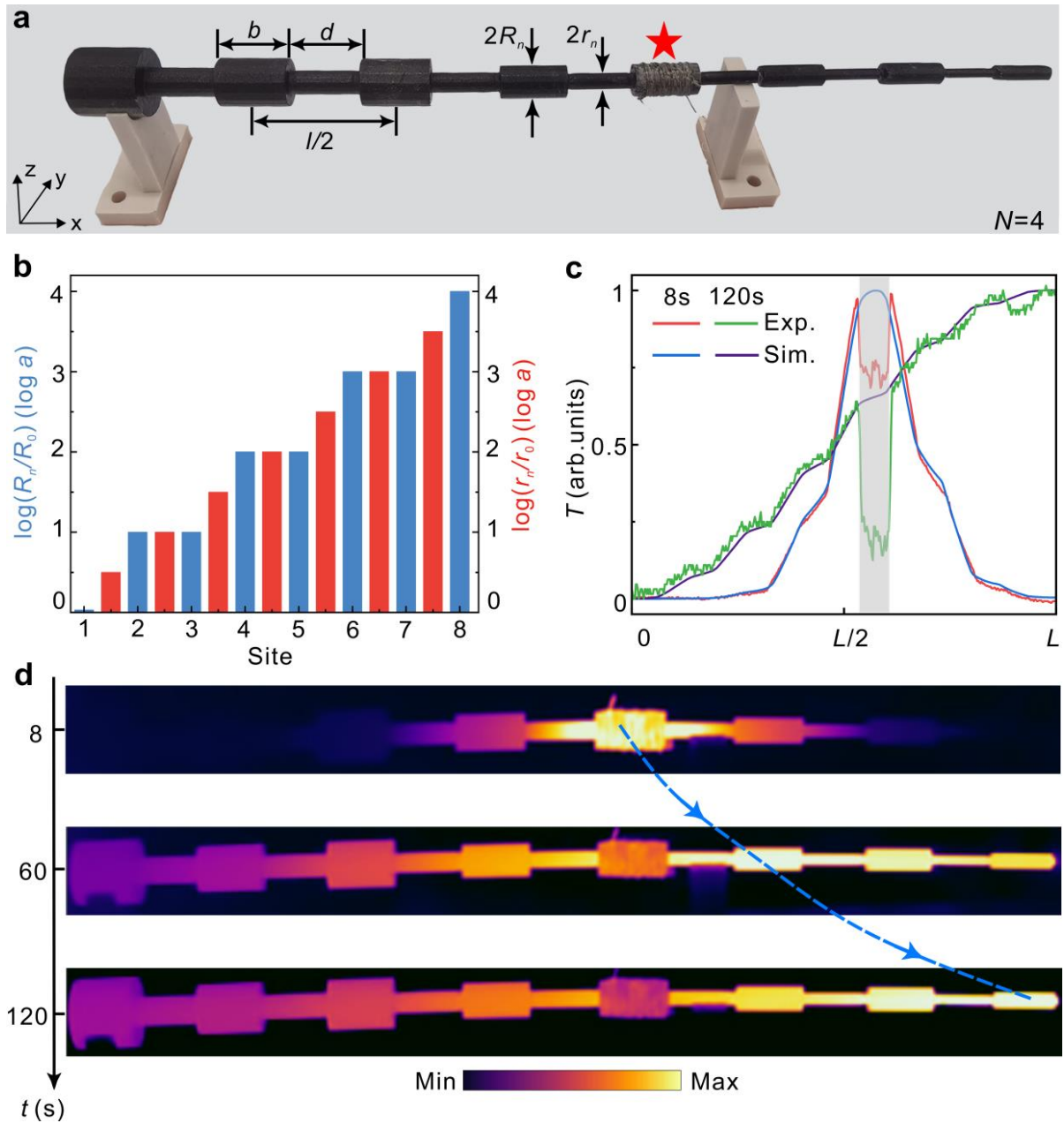

**Fig. 4 | Experimental observation of transient diffusive NHSE. a** Photograph of the 3D-printed samples with the structural parameters marked. The red asterisk shows the location of initial heat excitation. **b** Variation of $R_n$ (radii of site rods) and $r_n$ (radii of coupling sticks) at different sites. **c** The experimentally measured and numerically calculated temperature fields on the sample at 8 s and 120 s, respectively. The shaded bar covers the position of the initially heated region, where the twining resistance wire and silicone grease lead to the abrupt variation of the temperature distribution. **d** Thermographs of the heated sample at different times. Here the blue dashed arrows indicate the direction of temperature field evolution.

**Robustness of transient diffusive NHSE against defects**

Last but not least, we experimentally demonstrated the robustness of transient diffusive NHSE against the defects. Due to the nontrivial winding number, the transient diffusive NHSE is supposed to be topologically protected. In Fig. 5a, we show a defective thermal lattice sample with the defects introduced on three randomly chosen site rods. The detailed geometries of defects are shown by the close-up images as denoted by the blue arrows, which are either a pair of bulge loops or notch rings. The experimental procedure is the same as above in Fig. 4. We find that the experimental results are also consistent with the simulation ones. Both the starting and final states of the temperature field distributions (Fig. 5b) as well as the freeze-frames of thermographs (Fig. 5c) show that the transient diffusive NHSE is indeed defect-immune, where the observed temperature field is flowing towards the boundary as marked by the blue dashed arrow line in Fig. 5c.

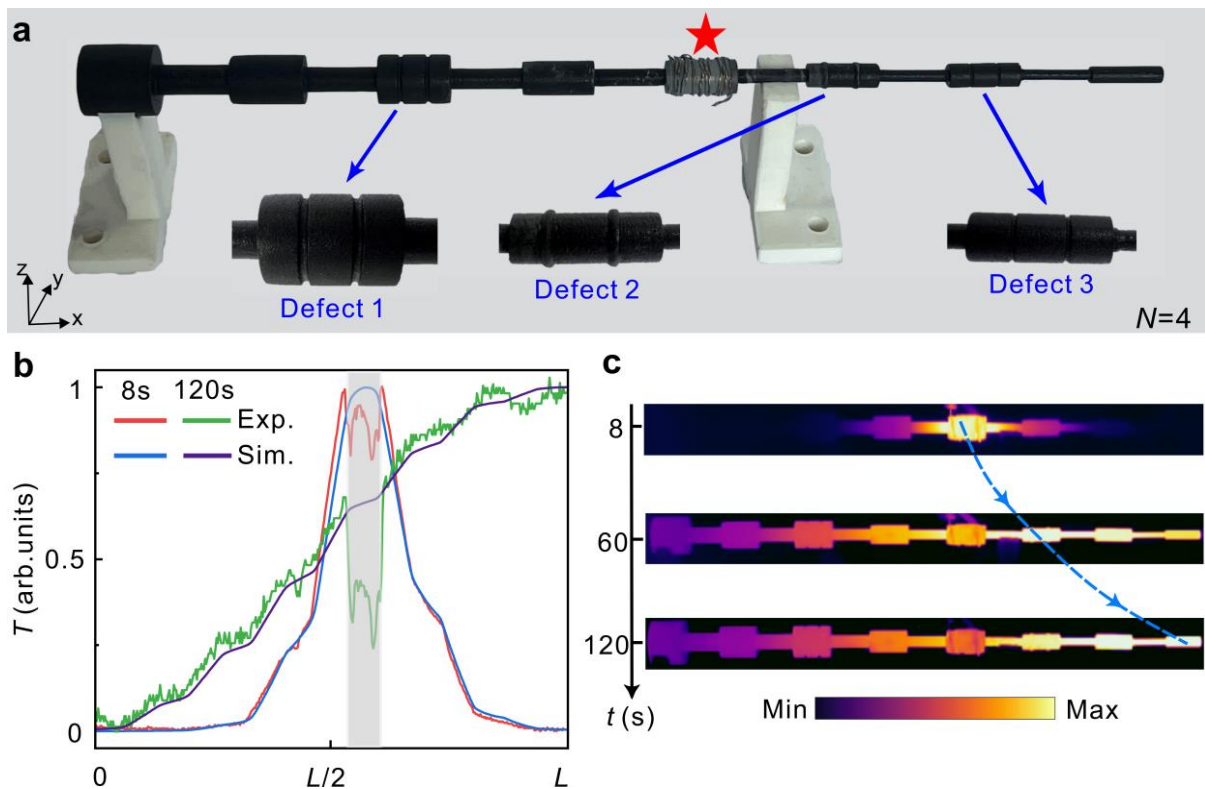

**Fig. 5 | Experimental demonstration of the robustness of diffusive NHSE against defects. a** Photograph of the thermal lattice with defects introduced at the randomly chosen site rods. **b**

The experimentally measured and numerically calculated temperature fields for the defective thermal lattice at 8 s, 120 s. **c** Thermographs of the heated sample at different times. The temperature field distributions were recorded at $t = 8$ s, $60$ s and $120$ s, respectively.

**Discussion**

In conclusion, our work provides the first experimental demonstration of the transient diffusive NHSE in a judiciously designed thermal metamaterial lattice. The diffusive NHSE is based on the asymmetric coupling in heat exchange process in a dimer cavity model, where the asymmetric temperature field distribution can be readily observed. We propose to aggregate the dimer unit cells to form a modified SSH thermal lattice. By introducing intra-cell asymmetric couplings into the thermal lattice, the anti-Hermitian diffusive system can be changed into a non-Hermitian one with a line gap, where the transient diffusive NHSE is predicted and verified in numerical simulations. Based on the design, we successfully fabricate the sample and experimentally observe the skin effect from temperature field evolutions and further prove that the skin effect of temperature field has strong immunity against defects. Our proposed approach is not just limited in the 1D case, but can be further extended into higher dimensions. Moreover, our work is expected to open an avenue for the non-Hermitian topology in diffusive regimes and benefit the design of functional thermo-devices with high sensitivity and robustness against deformation or damages.

**Methods**

**Numerical Simulations** All simulations calculating the eigenmodes and the transient processes were performed in the commercial finite element solver COMSOL Multiphysics 6.0. The 3D solid heat transfer module was chosen for the simulations in the article. The structure surface is set with the thermal insulation boundary condition. The volume specific heat capacity $\rho c_{p_0}$ of site rods and the thermal conductivity $\kappa_0$ of the coupling sticks are $\rho c_{p_0} = 2.43 \times 10^6$ J/m³ · K, $\kappa_0 = 180$ W/m · K. The ambient temperature is set to be 293.15 K, which also is the initial temperature of the unheated region. The temperture field distribution is extracted along an axial line of the cylindrical structure.

**Experimental Methods** We used the Aluminium alloy (AlSi$_{10}$Mg) for fabricating the thermal lattice, for which the experimental samples were fabricated by selective laser melting (SLM) technology of 3D metal printing. The manufacturing precision of SLM technology can reach ± 0.1 mm/100 mm. We placed the sample in a vacuum chamber and suspended it with plastic stands of poor thermal conductivity. The resistance wire interwining one site rod was connected to an adjustable power supply outside the vacuum chamber by a circuit for loading the heat source. At the beginning of experiment, the circuit was switched on and a large current passed through the resistance wire to generate a large amount of heat in a short time. Then we turned off the power supply and stopped heating for the following observation of temperature field evolution. The evolution of temperature field was recorded by using an infrared camera (Fotric 347).

**Data availability**

The data that support the findings of this study are available from the corresponding author upon reasonable request.

**Code availability**

The code utilized during the current study is available from the corresponding author on request.

**Acknowledgements**

The work is sponsored by the Key Research and Development Program of the Ministry of Science and Technology under Grant 2022YFA1405200, the National Natural Science Foundation of China (NNSFC) under Grants No. 92163123 and 52250191, the Key Research and Development Program of Zhejiang Province under Grant No.2022C01036, and the Fundamental Research Funds for the Central Universities.

**Author contributions**

Y.-K. Liu and P.-C. Cao contributed equally to this work. Y. Li, Y.-G. Peng, and X.-F. Zhu conceived the idea and supervised the project. Y.-K. Liu and P.-C. Cao developed the theory and designed the model. Y.-K. Liu, P.-C. Cao and Qiang-Kai-Lai Huang analyzed the diffusive SSH model. Y.-K. Liu and M. Qi performed the experiments. All authors contributed to the writing and discussions of the manuscript.